\documentclass[twocolumn,prl,showpacs]{revtex4}

\usepackage{amssymb}

\usepackage{epsfig}

\begin{document}

\title{Tracking heterogeneous dynamics during the $\alpha$-relaxation of a simple glass-former}

\author{Pinaki Chaudhuri $^{1}$, Srikanth Sastry $^{2}$ and Walter Kob $^{1}$ }

\affiliation{
$^1$ LCVN, UMR 5587, 
Universit\'e Montpellier II and CNRS, 
34095 Montpellier, France\\
$^2$ Jawaharlal Nehru Centre
for Advanced Scientific Research, Jakkur Campus, Bangalore 560064, India }

\date{}

\begin{abstract}
We study the relaxation process in a simple glass-former - the KA
lattice gas model. We show that, for this model, structural relaxation
is due to slow percolation of regions of co-operatively moving particles, 
which leads to heterogeneous dynamics of the system. We
find that the size distribution of these regions is given by a power-law
and that their formation is encoded in the initial structure of the
particles, with the memory of initial configuration increasingly
retained with increasing density.

\end{abstract}

\maketitle

Glass-forming systems are disordered materials whose relaxation dynamics
becomes extremely slow on decreasing the temperature or increasing
the density. Despite intensive research in this domain, a proper
understanding of these materials is still missing~\cite{deben,ediger}.
The observation, in experiments and simulations, of spatio-temporal
dynamical heterogeneity in glass-formers has been an important step
forward in elucidating the mechanism for relaxation processes and
currently considerable research is being focused on understanding the
lifetime and spatial extent of these dynamical heterogeneities.

A key question in this domain is the conundrum regarding whether or not
there is a causal link between structural properties of glass-forming
systems and their dynamical behavior.  It has recently been shown
\cite{harro} that the spatial heterogeneity in the propensity of
particles to move is correlated with the local environment of the
particles, characterized by the local Debye-Waller factor. However, this
influence seems to exist only over timescales which are much less than
the structural relaxation time~\cite{appi}.  It was also observed that
structural properties are indeed correlated with collective dynamical
fluctuations, but no quantitative analysis was made~\cite{ludorob}.
More recently, it has also been reported~\cite{harro1} that for a
supercooled liquid configuration, its localized low-frequency
normal modes correlate with the irreversible structural 
reorganization of its constituent particles.

In recent times, extensive studies of kinetically constrained models (KCM-s)
\cite{kcm}, which are one of the simplest models showing glassy dynamics,
have been carried out in order to understand their relaxation process.
These models, which are motivated by the hypothesis that the slow
dynamics in glass-formers is only due to geometrical constraints, show
heterogeneous dynamics similar to real glass-formers \cite{kcmdyn}.
In this Letter, we study one such KCM -
the Kob-Andersen (KA) lattice gas \cite{kob} in which particles are
allowed to move on a lattice following certain dynamical rules and which
at high densities shows signatures of apparently diverging relaxation
times~\cite{kob, franz, pitard}.  Recently it has been proven
analytically \cite{toni} and numerically \cite{ludofinite} that for this
model there exists no dynamical transition at finite density, $\rho$, and it was
argued \cite{toni} that eventually, due to the presence of migrating
macro-vacancies, the system relaxes, albeit extremely slowly. However,
from a practical point of view this model still is a good model for a
glass-forming systems. Using Monte Carlo simulations
we show that its structural relaxation is related to the
growth of mobile regions and that this process quickly slows down with
increasing $\rho$, resulting in the observed slow dynamics. We
also demonstrate that the formation of the mobile regions is directly
related to structural properties of the system.

We have studied the 3d version of the KA model: $N$ particles populate
a cubic lattice of size $L^3$ with the constraint that a lattice site can
be occupied by only one particle. All possible configurations have the
same energy and thus the same Boltzmann weight.  The imposed stochastic
dynamics consists of the following process: A randomly selected particle
can move to any one of the neighboring empty lattice site provided it
has $m$ or fewer occupied nearest neighbor site and that the target
empty site has $m + 1$ or fewer occupied nearest neighbor sites. A
choice of $m=3$ results in glassy dynamics for this model~\cite{kob}.
For efficient sampling of the configuration space at high $\rho=N/L^3$, we
have carried out event-driven Monte Carlo \cite{kob} simulations of the
model. Using periodic boundary conditions, we have investigated system
sizes $L=20$, 30, and 50, which avoid finite size effects, with
densities spanning from $\rho=0.65$ to $\rho=0.89$.

\begin{figure}[]
\vspace*{-5mm}
{\includegraphics[scale=0.65,angle=-90]{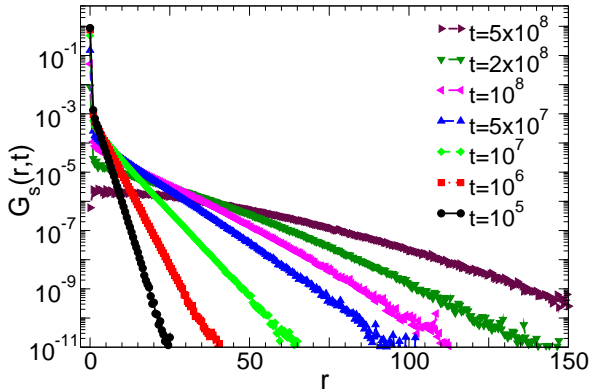}}
\vspace*{-6mm}
\caption{$G_s(r,t)$ for different times at $\rho=0.87$. The
$\alpha-$relaxation time at this density is $\tau_\alpha \approx
1.8\times10^7$.}
\label{gsrt87}
\vspace*{-5mm}
\end{figure}

Experiments and simulations in which the motion of single particles
were tracked have helped to demonstrate the existence of
heterogeneous dynamics in glassy systems~\cite{ediger}. E.g.,  by measuring
the self part of the van Hove function $G_s(r,t)$,
i.e. the distribution of particle displacements ($G_s (r,t)= \langle
\delta ( r - |{\bf r}_i(t) - {\bf r}_i(0)|) \rangle$, where ${\bf
r}_i(t)$ denotes the position of particle $i$ at time $t$), it has been
possible to demonstrate that the particles have varying mobilities.
In  Fig.~\ref{gsrt87}, we show $G_s(r,t)$ for different times $t$,
measured in units of Monte Carlo steps, at $\rho=0.87$.  For diffusive
motion, $G_s (r,t)$ is a Gaussian and we observe that for
the KA model, like other glass-formers \cite{szamel1}, this Gaussian
behavior is only observed at times ($t\approx{5\times10^8}$) that
are much larger than the structural relaxation time $\tau_\alpha$ (defined as the
time at which the self-intermediate scattering function has decayed to
$1/e$, which at $\rho=0.87$ is $\tau_\alpha \approx 1.8\times10^7$).
At intermediates times, we see that $G_s(r,t)$ has an exponential
tail, a signature of the presence of rare events in the dynamics of the
particles, similar to other glassy systems~\cite{pinaki}. Thus we can
conclude that while most of the particles remain frozen at their initial
positions (resulting in large values for $G(0,t)$), there is a small
population which is extremely mobile, i.e. the system has a very heterogeneous
dynamics.

\begin{figure}[]
\vspace*{-5mm}
\includegraphics[scale=0.13,angle=0]{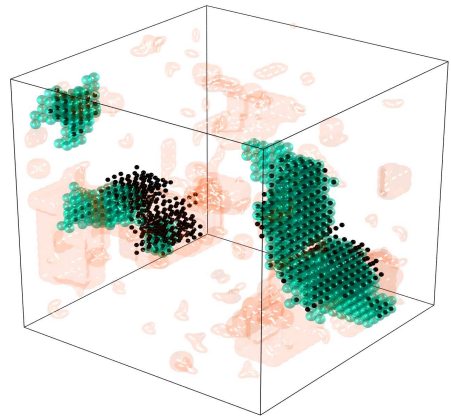}
\vspace*{-5mm}
\caption{Green and black points: Sites that have been visited by
the trajectory of two mobile particles. Red blobs: The mobility regions. $t=10^7$  and
$\rho=0.88$.}
\label{trajxyz}
\vspace*{-5mm}
\end{figure}

By tracking the mobile particles, i.e. particles which contribute to the
tail of $G_s(r,t)$, we see that at short times they explore a compact
region (a ``blob'') around their initial locations.  With increasing
$t$ these blobs slowly expand and coalesce with other blobs to form
a labyrinthine structure. This structure allows particles, which
were hitherto confined to one blob, to travel longer distances. In
Fig.~\ref{trajxyz} we have plotted the lattice sites (marked by green
and black spheres) visited by two such mobile particles at $\rho=0.88$
and $t=10^7$ (at this density, $\tau_\alpha \approx 4.4\times10^8$). It
can be clearly seen that the dynamics is spatially heterogeneous since the
trajectories consist of connected blobs. Note that initially each blob is
a region of {\it cooperative} motion since {\it all} the particles which
occupy these lattice points are found to be mobile. As the backbone is
formed, particles from one blob begin to explore other blobs. 
The relaxation of the system thus happens with the slow growth of this 
backbone, within which the particles can move relatively quickly.
This is demonstrated in Fig.~\ref{trajxyz} : indeed, the two trajectories 
overlap even though they originate
from two different lattice sites which initially did not belong to the
same blob.

Further insight into the spatial nature of the relaxation process
can be obtained by observing the so-called ``mobility regions''
\cite{pan,gar2,lawlor}: A lattice site is defined to be an ``active
site'' if either a particle or a vacancy has moved out of it during
the time of observation and the collection of these sites constitute
the mobility regions. Earlier studies of KCM-s
have shown that the active sites tend to cluster and act as seeds for
subsequent mobility \cite{pan,gar2,lawlor} and that it is possible to
extract from the mobility regions a lengthscale which increases with
density \cite{pan}. In Fig.~\ref{trajxyz} we have also included the
location of the active sites and we can see that these lattice sites
are indeed clustered in space (marked by the red blobby shapes) and
have a labyrinthine structure. It gives us an idea of the structure
of pathways, at high densities, available to the mobile particles for
exploration. We can clearly see that, on this time-scale, the two mobile
particles have only explored a part of the available volume and that the
geometry of the mobility regions and the blob-structure of the trajectories
are intimately connected to each other.

\begin{figure}[t]
\vspace*{-7mm}
\includegraphics[scale=0.30,angle=-90]{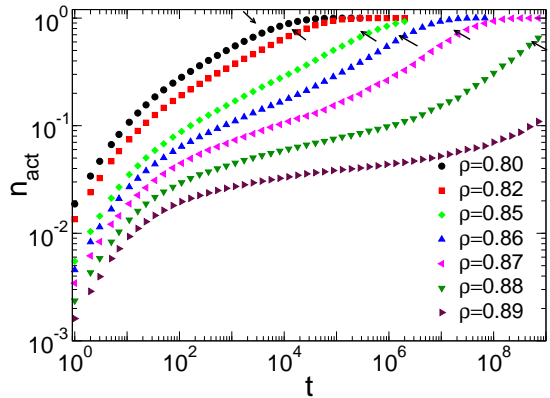}
\vspace*{-5mm}
\caption{Growth of number of active sites, $n_{act}$, as a function
of time, for different $\rho$. The arrows mark $\tau_\alpha$.}
\label{mob2}
\vspace{-5mm}
\end{figure}

The number density of active sites, $n_{act}(t)$, allows us to estimate
the volume accessible to the mobile particles and in Fig.~\ref{mob2}
we plot $n_{act}(t)$ for different $\rho$. We see that, at short
times, $n_{act}(t)$ increases quickly and we find $n_{act}(t)\sim
1-\exp(-t/\theta)$ with $\theta\approx5$, independent of $\rho$. This
regime corresponds to the initial growth of the blobs. Subsequently the
shape of $n_{act}(t)$ depends strongly on $\rho$.  For $\rho=0.80$ the
number of inactive sites, $1-n_{act}$, decays with a stretched exponential
tail, with a stretching exponent of around $0.6$, a functional form
that is found for all $\rho$. At even larger $\rho$, $n_{act}(t)$ shows
three regimes, with the second regime being a period of extremely slow
growth, almost logarithmic and thus similar to the coarsening process
in disordered media \cite{logs}. Note that at short $t$ the typical
distance between the blobs increases with $\rho$ and the growth of the
blobs slows down with increasing $\rho$, since it needs the presence of
active sites (which are rare at high density). This is the reason why
the increase of $n_{act}(t)$ at intermediate times becomes very slow with
increasing $\rho$. We also observe that at $\rho=0.89$, which is higher
than the density of $\rho_c=0.881$ at which an apparent divergence of
relaxation timescales was observed \cite{kob}, $n_{act}(t)$ is still an
increasing function, suggesting that the system will eventually relax
\cite{toni}. Note that although these three regimes are in qualitative 
agreement with the predictions of Ref.~\cite{toni}, there are important differences 
since, e.g., the diffusion of macro-vacancies discussed in Ref.~\cite{toni}
would lead to a linear growth of $n_{act}(t)$ at long times, a behaviour
which is not seen in Fig.~\ref{mob2}. This might be due to the fact that
the calculations presented in Ref.~\cite{toni} apply only at densities
that are extremely close to 1.0.

\begin{figure}[b]
\vspace*{-7mm}
\includegraphics[scale=0.70,angle=-90]{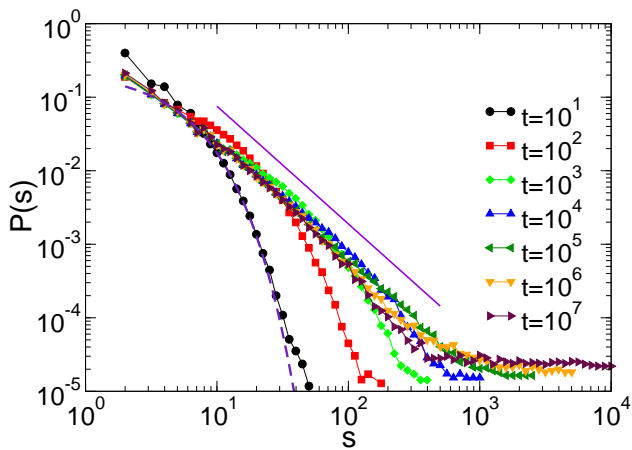}
\vspace*{-9mm}
\caption{Distribution of cluster size, $P(s)$, where $s$ is the size of
cluster, for different times at  $\rho=0.88$. Also shown (dashed
line) is an exponential fit to $P(s)$ at $t=10$ and (solid line) 
the function ${s^{-1.6}}$ to highlight the emergence of power-law
behavior of $P(s)$ at intermediate times.}
\label{ps88}
\end{figure}

In order to characterize the geometry of the growing clusters of active
sites we have calculated $P(s)$, the distribution of clusters
that have exactly size $s$. In Fig.~\ref{ps88}, we have plotted $P(s)$
for $\rho=0.88$ and different times. At short times, $t=10$, $P(s)$
has an exponential shape. This corresponds to the initial geometry
of active sites at few random locations when particles explore their
neighborhood. With increasing time $P(s)$ quickly transforms into a
power-law, $P(s)\sim{s^{-\nu}}$, with an exponent $\nu\approx{1.6}$. This
indicates that the growth process is different from random percolation
for which $\nu=2.2$. This difference is likely related to the fact that
there is a wide variation in the size of the cooperatively rearranging
regions seen at short times. At even later times, the largest
cluster starts growing and the tail in $P(s)$ shifts to larger and larger
sizes, until the entire space is filled up. Note that the observed {\it largest} cluster
dependens on system-size and thus the tail of $P(s)$ is affected by finite-size
effects.

\begin{figure}[t]
\vspace*{-7mm}
\includegraphics[scale=0.6,angle=-90]{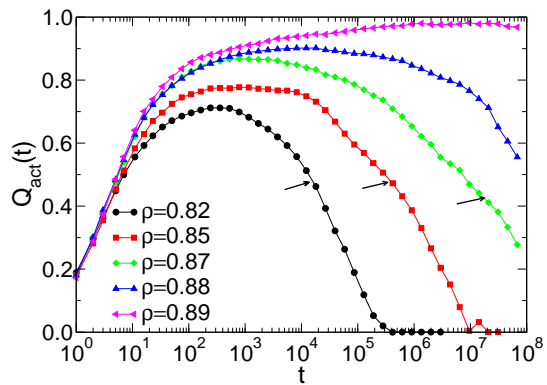}
\vspace*{-2mm}
\caption{Variation of $Q_{act}(t)$ with time, for different $\rho$.
The arrows mark  $\tau_\alpha$.}
\label{actovtraj}
\vspace{-6mm}
\end{figure}

Finally we investigate to what extent the dynamics is encoded in the
structure. We have seen that structural relaxation is correlated with
the development of the mobile regions. Therefore it is interesting to
check how the properties of these regions depend on the trajectories
that start from the same initial configuration. For this, we check
along several such trajectories how different are the configurations 
of active sites formed at the same observation time. To quantify
that, we define the overlap function $Q_{act}(t)=(\langle{q^{\alpha\beta}(t)}
\rangle_{ic}-n^2_{act}(t))/(n_{act}(t)-n^2_{act}(t))$, where
$q^{\alpha\beta}(t)=L^{-3}\sum_{i}n_i^{\alpha}(t)n_i^{\beta}(t)$, with
$n_i(t)=1$ if site $i$ is active at time $t$ and $n_i(t)=0$ otherwise.
$\langle . \rangle_{ic}$ is the average over the isoconfigurational
ensemble \cite{harro}, i.e. the ensemble of all possible trajectories
starting from the same configuration of which $\alpha$ and $\beta$ are
two different members. Defined in this way, $Q_{act}(t)=1$ if at time
$t$ the configuration of active sites for two different trajectories are
exactly the same and $Q_{act}(t)=0$ if the two configurations are totally
different, apart from the trivial statistical overlap. Thus $Q_{act}$
is a direct measure for the influence of the initial structure on the
mobility regions.  

In Fig.~\ref{actovtraj}, we show the time dependence
of $Q_{act}(t)$ for different $\rho$. At short times $Q_{act}(t)$ is
independent of $\rho$ and its value is small since the mobile particles
can find random directions to explore, which result in different
configurations of active sites and hence a small $Q_{act}$. Subsequently
$Q_{act}(t)$ has a peak at a time which approximately corresponds to
the time at which $n_{act}(t)$ enters the final regime of growth, see
Fig.~\ref{mob2}, and which is much smaller than $\tau_\alpha$ (marked
by arrows). For large value of $\rho$ the height of this peak is close
to unity and it becomes extremely broad so that $Q_{act}(t)$ is quite
large even at $\tau_\alpha$.  This shows that the configuration of active
sites is, in a broad time window, basically independent of the trajectory
and thus encoded in the initial structure, even for times of the order
of $\tau_\alpha$ (but depends of course on the initial configuration).
Note that these results seem to be in contradiction to the claims made in
Ref.~\cite{appi} since there it was argued that the structure influences
the dynamics only on time scales much shorter than $\tau_\alpha$. However,
in that work the authors considered this influence on the level of {\it
individual} particles whereas the overlap $Q_{act}(t)$ considered here is
a {\it collective} quantity. Therefore there is not necessarily a contradiction.

\begin{figure}[t]
\vspace*{-5mm}
\includegraphics[scale=0.6,angle=-90]{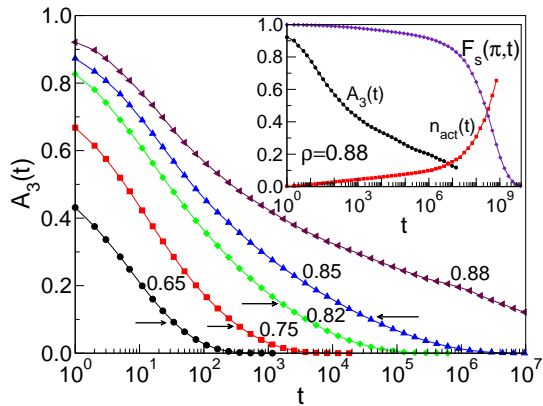}
\vspace*{-5mm}
\caption{The time evolution of overlap, $A_3(t)$, for different $\rho$.
The arrows mark $\tau_\alpha$.  Inset:
Comparison of different temporal functions at $\rho=0.88$.}
\label{overlap}
\vspace{-3mm}
\end{figure}

What structural property determines mobility? It is obvious that for a
site to be active, it needs empty space in its neighborhood. Earlier
work on lattice models has shown that the high propensity sites are
located near clusters of empty sites~\cite{gar2}. In order to investigate
the relation between structure and mobility we define another overlap
function. A lattice site is considered to be a {\it generalized vacancy of
type $k$} if the site and its six nearest neighbors contains a total of at
least $k$ holes. Then we calculate, as a function of time, the overlap of
the active sites with these generalized vacancies. The quantity we measure
is $A_k(t)=\sum_i{a_i^kn_i(t)}/\sum_i{n_i(t)}-\tilde{a}^k$. Here, $a_i^k=1$
if at $t=0$ the site $i$ is a generalized vacancy of type $k$ and $0$
otherwise. Thus, $A_k(t)$ is the probability that a generalized vacancy
of type $k$ in the configuration at $t=0$ is an active site at time
$t$, with the trivial overlap $\tilde{a}^k$ (the density of generalized
vacancies of type $k$ in the configuration at $t=0$) subtracted. Such
a quantity enables us to have a good measure of the correlation
between structure and dynamics. For low density, i.e. $\rho=0.65$,
$A_3(t)$ has a fast decay, see Fig.~\ref{overlap}. (Other value of
$k$ have a qualitatively similar behavior.) With increasing $\rho$, the
characteristic time-scale for the decay increases with the tail becoming
stretched in shape, the underlying slowing down of the relaxation process
resulting in the retention of the memory for longer time. In the inset
of Fig.~\ref{overlap}, we have plotted different correlation functions
at $\rho=0.88$: The overlap function $A_3(t)$, the self-intermediate
scattering function $F_s(\pi,t)$ and the fraction of active sites
$n_{act}(t)$.  At this density, the $\alpha$-relaxation time $\tau_\alpha
\approx 4.4\times{10^8}$. For this time-scale, the fraction of active
sites is $n_{act}(\tau_\alpha)\sim{0.50}$. However, if one extrapolates
$A_3(t)$ to these time-scales, the overlap is small. Therefore,
although measurement of $Q_{act}(t)$ showed that the configuration of
active sites, at $t\approx\tau_{\alpha}$, is significantly determined
by initial structure, the use of generalized vacancies does not fully
demonstrate this strong dependence. Hence, a better characterization of
initial structure is necessary for improving prediction for formation
of mobile regions.

In conclusion, we have shown that, for the KA lattice gas, the
$\alpha-$relaxation occurs via the percolation of mobile regions in
which particles move cooperatively. These regions are encoded in the
initial structure of the system with a memory time that is on the order
of $\tau_\alpha$. We emphasize, however, that the initial structure
does not necessarily determine the trajectory of an {\it individual}
particle but only the location and the shape of the regions in which
cooperative dynamics is observed.

We thank L. Berthier and G. Biroli for useful discussions and acknowledge
CEFIPRA Project 3004-1 and ANR Grant TSANET for financial support.


\vspace{-3mm}

\begin{thebibliography}{10}
\bibitem{deben} 
P. G. Debenedetti {\it et al},
Nature {\bf 410}, 259 (2001).

\bibitem{ediger} 
M. A. Ediger, 
Annu. Rev. Phys. Chem. {\bf 51}, 99 (2000).

\bibitem{harro}
A. Widmer-Cooper {\it et al}, 
Phys. Rev. Lett. {\bf 96}, 185701 (2006).

\bibitem{appi}
G. A. Appignanesi {\it et al},
Phys. Rev. Lett. {\bf 96}, 237803 (2006).

\bibitem{ludorob} 
L. Berthier {\it et al}, 
Phys. Rev. E {\bf 76}, 041509 (2007).

\bibitem{harro1}
A. Widmer-Cooper {\it et al},
Nature Physics {\bf 4}, 711 (2008).

\bibitem{kcm} 
F. Ritort and P. Sollich, 
Adv. Phys. {\bf 52}, 219 (2003).

\bibitem{kcmdyn}
M. Foley {\it et al}, 
J. Chem. Phys., {\bf 98}, 5069  (1993); 
%
J.P. Garrahan {\it et al}, 
Phys. Rev. Lett. {\bf 89}, 035704 (2002);
%
J. J\"{a}ckle, 
J. Phys.: Condens. Matt,  {\bf 14}, 1423 (2002);
%
L. Berthier {\it et al}, 
Europhys. Lett. {\bf 69}, 320 (2005).

\bibitem{kob} 
W. Kob {\it et al}, 
Phys. Rev. E. {\bf 48} 4364 (1993).

\bibitem{franz} 
S. Franz {\it et al},
Phys. Rev. E {\bf 65} 021506 (2002).

\bibitem{pitard} 
E. Marinari {\it et al},
Europhys. Lett. {\bf 69}, 235 (2005).

\bibitem{toni}
C. Toninelli {\it et al},
Phys. Rev. Lett. {\bf92}, 185504 (2004); 
J. Stat. Phys. {\bf 120}, 167 (2005). 

\bibitem{ludofinite} 
L. Berthier, 
Phys. Rev. Lett. {\bf 91}, 055701 (2003).

\bibitem{szamel1} 
G. Szamel {\it et al},
Phys. Rev. E {\bf 73}, 011504 (2006).

\bibitem{pinaki}
P. Chaudhuri {\it et al},
Phys. Rev. Lett. {\bf 99}, 060604 (2007).

\bibitem{pan} 
A.~C. Pan {\it et al},
Phys. Rev. E. {\bf 72}, 041106 (2005).

\bibitem{gar2}
L.~O. Hedges {\it et al},
J. Phys.: Cond. Matt. {\bf 19}, 205124 (2007).

\bibitem{lawlor}
A. Lawlor {\it et al},
Phys. Rev. E {\bf 72}, 021401 (2005).

\bibitem{logs} 
A.~J. Bray, 
Adv. in Phys., {\bf 43} 357 (1994).

\end{thebibliography}
\end{document}